\documentclass[12pt]{iopart}
\usepackage{iopams}  
\usepackage{bm}
\usepackage{graphicx}

\begin{document}

\title[Photon emission from macroscopic currents]{
Photon emission from macroscopic currents}

\author{Rainer Dick}

\address{Department of Physics and Engineering Physics, 
University of Saskatchewan, 116 Science Place, Saskatoon, Canada SK S7N 5E2}

\begin{abstract}  
 Coherent states are a well-established tool of quantum optics to describe
  electromagnetic waves in terms of photons. However, they do not describe
  the near-field regime of radiation sources.
  Instead, we generically use classical
  solutions of Maxwell's equations to describe radiation in the near-field regime.
  The classical solutions provide linear relations between
  currents and emitted electromagnetic fields, whereas evolution of states
  at the quantum level proceeds through unitary time evolution operators
  involving photon operators. This begs questions how the classical
  radiation equations relate to unitary quantum evolution, and how
  we can describe macroscopic fields from antennas or magnetic coils
  in terms of elementary photons. The present paper answers both questions
  through the construction of generalized Glauber states for radiation
  emitters.
\end{abstract}


%
\vspace{2pc}
\section{Introduction}\label{sec:intro}

After their introduction into the realm of quantum
optics \cite{rf:glauber1,rf:sudarshan,rf:glauber},
coherent states became the universal tool to describe electromagnetic
waves in terms of photons \cite{rf:scully,rf:loudon,rf:knight,rf:fox,rf:agarwal}.
They become ever more relevant for comparing classical versus quantum behavior of light
\cite{rf:chen,rf:kleinbeck,rf:vinu,rf:lange}, for investigations of imaging when quantum behavior
of electromagnetic waves becomes relevant \cite{rf:villegas}, and for quantum information and
quantum computing \cite{rf:qi,rf:labay}.

Coherent states are usually employed with a focus on single photon modes, although
Glauber had already formulated them for superpositions of arbitrary many discrete
$\bm{k}$-vectors\footnote{We use $\zeta(\bm{k})$ or $\zeta(k)$ instead of
  Glauber's $\alpha_{\bm{k}}$ or Loudon's $\alpha(\omega)=\alpha(ck)$
  for the amplitudes, because we also include summations over polarization
  indices $\alpha$.}
\cite{rf:glauber},
\begin{eqnarray}\nonumber
\bm{|}\bm{\zeta}\bm{\rangle}&=&\exp\!\left(\sum_{\bm{k},\alpha}
\!\left[\zeta_\alpha(\bm{k})a_{\alpha}^+(\bm{k})
-\zeta_{\alpha}^+(\bm{k})a_\alpha(\bm{k})\right]\!\right)\!\bm{|}0\bm{\rangle}
\\ \label{eq:glauber}
&=&\exp\!\left(\sum_{\bm{k},\alpha}
\!\left(\zeta_\alpha(\bm{k})a_{\alpha}^+(\bm{k})
-\frac{1}{2}\!\left|\zeta_{\alpha}(\bm{k})\right|^2\right)\!\right)\!\bm{|}0\bm{\rangle},
\end{eqnarray}
while Loudon also discussed continuous superpositions
of frequencies for fixed direction $\hat{\bm{k}}$ \cite{rf:loudon},
\begin{eqnarray}\nonumber
\bm{|}\bm{\zeta}\bm{\rangle}&=&\exp\!\left(\int_0^\infty\!\!dk
\sum_{\alpha=1}^2\!\left[\zeta_\alpha(k)a_{\alpha}^+(k)
-\zeta_{\alpha}^+(k)a_\alpha(k)\right]\!\right)\!\bm{|}0\bm{\rangle}
\\  \label{eq:loudon}
&=&\exp\!\left(\int_0^\infty\!\!dk
\sum_{\alpha=1}^2\left[\zeta_\alpha(k)a_{\alpha}^+(k)-\frac{1}{2}
\left|\zeta_{\alpha}(k)\right|^2\right]\right)\!\bm{|}0\bm{\rangle}.
\end{eqnarray}
The photon operators in Eq.~(\ref{eq:loudon}) are normalized in the $k$-scale,
 $[a_\alpha(k),a_{\beta}^+(k')]=\delta_{\alpha\beta}\delta(k-k')$.

The generalization of Eq.~(\ref{eq:glauber}) to continuous photon modes is
straightforward, 
\begin{eqnarray}\nonumber
\bm{|}\bm{\zeta}\bm{\rangle}&=&\exp\!\left(\int\!d^3\bm{k}
\sum_{\alpha=1}^2\left[\zeta_\alpha(\bm{k})a_{\alpha}^+(\bm{k})
-\zeta_{\alpha}^+(\bm{k})a_\alpha(\bm{k})\right]\!\right)\!\bm{|}0\bm{\rangle}
\\ \label{eq:zetastate}
&=&\exp\!\left(\int\!d^3\bm{k}
\sum_{\alpha=1}^2\left[\zeta_\alpha(\bm{k})a_{\alpha}^+(\bm{k})
-\frac{1}{2}\left|\zeta_{\alpha}(\bm{k})\right|^2\right]\right)\!\bm{|}0\bm{\rangle}.
\end{eqnarray}
In the following, we define the vectors
\begin{equation}
\bm{\zeta}(\bm{k})=\sum_{\alpha=1}^2\zeta_\alpha(\bm{k})\bm{\epsilon}_\alpha(\bm{k}),
\end{equation}
and the operators
\begin{equation}
\bm{a}(\bm{k})=\sum_{\alpha=1}^2 a_\alpha(\bm{k})\bm{\epsilon}_\alpha(\bm{k}),
\end{equation}
where $\bm{k}\cdot\bm{\epsilon}_\alpha(\bm{k})=0$
and $\bm{\epsilon}^+_\alpha(\bm{k})\cdot\bm{\epsilon}_\beta(\bm{k})=\delta_{\alpha\beta}$.
The canonical commutation relations
\begin{equation}
[a_\alpha(\bm{k}),a^+_\beta(\bm{k}')]=\delta_{\alpha\beta}\delta(\bm{k}-\bm{k}')
\end{equation}
imply
\begin{eqnarray}\nonumber
  [\bm{a}(\bm{k})\stackrel{\otimes}{,}\bm{a}^+(\bm{k}')]&\equiv&
  \bm{a}(\bm{k})\otimes\bm{a}^+(\bm{k}')-\bm{a}^+(\bm{k}')\otimes\bm{a}(\bm{k})
  \\
  &=&\!\left(\underline{1}-\hat{\bm{k}}\otimes\hat{\bm{k}}\right)\!\delta(\bm{k}-\bm{k}').
\end{eqnarray}

The states (\ref{eq:glauber},\ref{eq:loudon},\ref{eq:zetastate}) are
time-independent states in the Heisenberg picture, and they yield
electromagnetic waves through expectation values of the corresponding
operators, e.g.
\begin{eqnarray}\nonumber
\bm{\mathcal{E}}_\perp(\bm{x},t)&=&\bm{\langle}\bm{\zeta}\bm{|}\bm{E}_\perp(\bm{x},t)
\bm{|}\bm{\zeta}\bm{\rangle}
\\ \nonumber
&=&\mathrm{i}\sqrt{\frac{\hbar\mu_0 c^3}{(2\pi)^3}}
\int\!d^3\bm{k}\,\sqrt{\frac{k}{2}}
\Big(\bm{\zeta}(\bm{k})\exp[\mathrm{i}(\bm{k}\cdot\bm{x}-ckt)]
\\ \label{eq:classicalE}
&&-\,\bm{\zeta}^+(\bm{k})\exp[-\,\mathrm{i}(\bm{k}\cdot\bm{x}-ckt)]
\Big),
\end{eqnarray}
\begin{eqnarray}\nonumber
\bm{\mathcal{B}}(\bm{x},t)&=&\bm{\langle}\bm{\zeta}\bm{|}\bm{B}(\bm{x},t)
\bm{|}\bm{\zeta}\bm{\rangle}
\\ \nonumber
&=&\mathrm{i}\sqrt{\frac{\hbar\mu_0 c}{(2\pi)^3}}
\int\!\frac{d^3\bm{k}}{\sqrt{2k}}\,
\bm{k}\times
\Big(\bm{\zeta}(\bm{k})\exp[\mathrm{i}(\bm{k}\cdot\bm{x}-ckt)]
\\ \label{eq:classicalB}
&&-\,\bm{\zeta}^+(\bm{k})
\exp[-\,\mathrm{i}(\bm{k}\cdot\bm{x}-ckt)]
\Big).
\end{eqnarray}
The electromagnetic field operators in the Heisenberg picture
(in SI units and in Coulomb gauge) are \cite{rf:rdqm3ed}
\begin{eqnarray}\nonumber
\bm{A}(\bm{x},t)&=&\sqrt{\frac{\hbar\mu_0 c}{(2\pi)^3}}
\int\!\frac{d^3\bm{k}}{\sqrt{2k}}\,
\Big(\bm{a}(\bm{k})
\exp[\mathrm{i}(\bm{k}\cdot\bm{x}-ckt)]
\\ \label{eq:Amode}
&&+\,\bm{a}^+(\bm{k})
\exp[-\,\mathrm{i}(\bm{k}\cdot\bm{x}-ckt)]\Big),
\end{eqnarray}
and for the fields
\begin{eqnarray}\nonumber
\bm{E}_\perp(\bm{x},t)&=&-\,\frac{\partial}{\partial t}\bm{A}(\bm{x},t)
\\ \nonumber
&=&\mathrm{i}\sqrt{\frac{\hbar\mu_0 c^3}{(2\pi)^3}}
\int\!d^3\bm{k}\,\sqrt{\frac{k}{2}}\,
\Big(\bm{a}(\bm{k})\exp[\mathrm{i}(\bm{k}\cdot\bm{x}-ckt)]
\\ \label{eq:Emode}
&&-\,\bm{a}^+(\bm{k})
\exp[-\,\mathrm{i}(\bm{k}\cdot\bm{x}-ckt)]\Big),
\end{eqnarray}
\begin{eqnarray}\nonumber
\bm{B}(\bm{x},t)&=&\bm{\nabla}\times\bm{A}(\bm{x},t)
\\ \nonumber
&=&\mathrm{i}\sqrt{\frac{\hbar\mu_0 c}{(2\pi)^3}}
\int\!\frac{d^3\bm{k}}{\sqrt{2k}}\,\bm{k}\times
\Big(\bm{a}(\bm{k})
\exp[\mathrm{i}(\bm{k}\cdot\bm{x}-ckt)]
\\ \label{eq:Bmode}
&&-\,\bm{a}^+(\bm{k})
\exp[-\,\mathrm{i}(\bm{k}\cdot\bm{x}-ckt)]
\Big).
\end{eqnarray}

The Hamiltonian is (we implicitly assume normal ordering)
\begin{eqnarray}\nonumber
  H_\gamma&=&\int\!d^3\bm{x}\left(\frac{\epsilon_0}{2}\bm{E}^2_\perp(\bm{x},t)+\frac{1}{2\mu_0}
  \bm{B}^2(\bm{x},t)\right)
  \\ \label{eq:Hgamma}
  &=&\int\!d^3\bm{k}\,\hbar ck\,\bm{a}^+(\bm{k})\cdot\bm{a}(\bm{k}).
\end{eqnarray}
The Schr\"odinger picture states are therefore
\begin{eqnarray}\nonumber
  \bm{|}\bm{\zeta}(t)\bm{\rangle}&=&\exp\!\left(-\,\mathrm{i}H_\gamma t/\hbar\right)\bm{|}
  \bm{\zeta}\bm{\rangle}
  \\ \nonumber
  &=&\exp\!\left(\int\!d^3\bm{k}\,\bm{\zeta}(\bm{k})\cdot\bm{a}^+(\bm{k})\exp(-\,\mathrm{i}ckt)
  \right.
  \\ \label{eq:zetastateS}
  &&-\left.\int\!d^3\bm{k}\,\bm{\zeta}^+(\bm{k})\cdot\bm{a}(\bm{k})
  \exp(\mathrm{i}ckt)\right)\!\bm{|}0\bm{\rangle}.
\end{eqnarray}
and we get, e.g., the electric field (\ref{eq:classicalE}) in the form
\begin{eqnarray}\nonumber
\bm{\mathcal{E}}_\perp(\bm{x},t)&=&\bm{\langle}\bm{\zeta}(t)\bm{|}\bm{E}_\perp(\bm{x})
\bm{|}\bm{\zeta}(t)\bm{\rangle}.
\end{eqnarray}

The limitation of the multi-mode
representations (\ref{eq:glauber},\ref{eq:loudon},\ref{eq:zetastate}) to free
electromagnetic waves was not an impediment for quantum optics, but it is not
satisfactory from a conceptual perspective, because we should also be able to
construct multi-mode decompositions of electromagnetic waves in terms of photons
near their generating currents. We will address this problem in Sec.~\ref{sec:rdstates}
and focus on the application to static magnetic fields from stationary current
densities $\bm{j}(\bm{x})$ in Sec.~\ref{sec:magnet}. We will find generalizations
of Eq.~(\ref{eq:zetastate}) to time-dependent
amplitudes $\zeta_\alpha(\bm{k})\to\xi_\alpha(\bm{k},t)$ in the Dirac interaction
picture, such that the electromagnetic fields from any current distribution
$\bm{j}(\bm{x},t)$ can be written as interaction-picture expectation values,
e.g.~$\bm{\mathcal{B}}(\bm{x},t)=\bm{\langle}\bm{\xi}(t)\bm{|}\bm{B}(\bm{x},t)
\bm{|}\bm{\xi}(t)\bm{\rangle}$,
where the parameters $\xi_\alpha(\bm{k},t)$
are linear functionals of the current densities. The relations then trivially allow
for the calculation of $\bm{k}$-space photon densities $n(\bm{k})$ and of total
photon numbers in electromagnetic fields from general current distributions.

\section{Generalized Glauber states for radiation from currents}\label{sec:rdstates}

Glauber was the first to point out
that multi-mode coherent states reproduce classical electromagnetic waves as expectation
values of the corresponding field operators \cite{rf:glauber}, and this is the aspect
that we are particularly interested in.
We will therefore denote the states as Glauber states for the purposes of generalizing
the multi-mode representation (\ref{eq:zetastate}) to near-field and static fields.

The notation $\bm{E}_\perp(\bm{x},t)$ for the transverse photon contribution (\ref{eq:Emode})
to the electric field reminds us that there is also the longitudinal electric field
operator,
\begin{equation}
\bm{E}_\|(\bm{x},t)=-\,\bm{\nabla}\Phi(\bm{x},t).
\end{equation}
In Coulomb gauge, this involves the charge density operators for the charged fields,
e.g.~for electrons
\begin{eqnarray}\nonumber
  \Phi(\bm{x},t)&=&\int\!d^3\bm{x}'\,
  \frac{\varrho(\bm{x}',t)}{4\pi\epsilon_0|\bm{x}-\bm{x}'|}
  \\
  &=&-\int\!d^3\bm{x}'\sum_{s=-1/2}^{1/2}
  \frac{e\psi_s^+(\bm{x}',t)\psi_s(\bm{x}',t)}{4\pi\epsilon_0|\bm{x}-\bm{x}'|},
\end{eqnarray}
where $s$ is the spin projection label.
However, the states (\ref{eq:glauber},\ref{eq:loudon},\ref{eq:zetastate}) assume that the
modes of the electromagnetic field are solutions of the free wave equation,
which is equivalent to the assumption that there are no
transverse currents
\begin{eqnarray}\label{eq:jperp}
  \bm{j}_\perp(\bm{x},t)&=&\int\!d^3\bm{x}'\,\underline{\delta}^{\perp}(\bm{x}-\bm{x}')\cdot
  \bm{j}(\bm{x}',t),
\end{eqnarray}
because in Coulomb gauge, Amp\`{e}re's law takes the form
\begin{equation}\label{eq:ampere}
  \left(\frac{1}{c^2}\frac{\partial^2}{\partial t^2}-\Delta\right)\!\bm{A}(\bm{x},t)
  =\mu_0\bm{j}_\perp(\bm{x},t).
\end{equation}
The projector in Eq.~(\ref{eq:jperp}) is the transverse $\delta$-function,
\begin{equation}
  \underline{\delta}^{\perp}(\bm{x})=\frac{1}{(2\pi)^3}\int\!d^3\bm{k}\left(
  \underline{1}-\hat{\bm{k}}\otimes\hat{\bm{k}}\right)\exp(\mathrm{i}\bm{k}\cdot\bm{x}).
\end{equation}

The states (\ref{eq:glauber},\ref{eq:loudon},\ref{eq:zetastate}) are therefore suitable for
the description of electromagnetic fields in the far-field region, far away from
transverse currents,
and in these cases they represent the asymptotic form of the Heisenberg-picture states.
However, they are not suitable in the near-field region. Furthermore,
the photonic resolution of static fields appears puzzling, since photons always
oscillate with frequency $ck>0$ and move with velocity $\bm{c}=c\hat{\bm{k}}$.

On the other hand, the operator equations
(\ref{eq:Amode},\ref{eq:Emode},\ref{eq:Bmode}) remain valid in the presence of sources,
in the sense that they
represent the photon quantum fields in the Dirac interaction picture, e.g.
\begin{equation}
  \bm{A}(\bm{x},t)=\exp\!\left(\mathrm{i}H_\gamma t/\hbar\right)\bm{A}(\bm{x})
  \exp\!\left(-\,\mathrm{i}H_\gamma t/\hbar\right).
\end{equation}
This helps us to generalize the state (\ref{eq:zetastate}) to a time-dependent state
$\bm{|}\bm{\xi}(t)\bm{\rangle}$ in the interaction picture such that we can reproduce
arbitrary electromagnetic fields with classical vector
potential $\bm{\mathcal{A}}(\bm{x},t)$ through expectation values in the
interaction picture,
\begin{eqnarray}  \nonumber
  \bm{\mathcal{A}}(\bm{x},t)&=&\bm{\langle}\bm{\xi}(t)\bm{|}\bm{A}(\bm{x},t)
  \bm{|}\bm{\xi}(t)\bm{\rangle}
  \\ \nonumber
  &=&\sqrt{\frac{\hbar\mu_0 c}{(2\pi)^3}}
\int\!\frac{d^3\bm{k}}{\sqrt{2k}}\,
\Big(\bm{\xi}(\bm{k},t)
\exp[\mathrm{i}(\bm{k}\cdot\bm{x}-ckt)]
\\ \label{eq:fixxi1}
&&+\,\bm{\xi}^+(\bm{k},t)
\exp[-\,\mathrm{i}(\bm{k}\cdot\bm{x}-ckt)]\Big).
\end{eqnarray}

 For the construction of the state
\begin{eqnarray} \label{eq:xistate1}
  \bm{|}\bm{\xi}(t)\bm{\rangle}
  &=&\exp\!\left(\int\!d^3\bm{k}\left[\bm{\xi}(\bm{k},t)\cdot\bm{a}^+(\bm{k})
  -\bm{\xi}^+(\bm{k},t)\cdot\bm{a}(\bm{k})\right]
  \right)\!\bm{|}0\bm{\rangle},
\end{eqnarray}
we use that the classical solution of Eq.~(\ref{eq:ampere})
is given in terms of the retarded Green's function,
\begin{equation}\label{eq:Gret1}
G(\bm{x},t)=\frac{c}{4\pi|\bm{x}|}\delta(ct-|\bm{x}|),
\end{equation}
\begin{eqnarray} \label{eq:solvamp1}
  \bm{\mathcal{A}}(\bm{x},t)&=&\mu_0\int\!d^3\bm{x}'\int_{-\infty}^\infty\!dt'\,
  G(\bm{x}-\bm{x}',t-t')
  \bm{j}_\perp(\bm{x}',t').
\end{eqnarray}
It is convenient for us to use the Fourier representation of the retarded Green's function,
\begin{eqnarray}\label{eq:Gret2}
  G(\bm{x},t)&=&\frac{1}{(2\pi)^3}\int\!d^3\bm{k}\,\exp(\mathrm{i}\bm{k}\cdot\bm{x})
  G(\bm{k},t),
\end{eqnarray}
with ($\Theta(t)$ is the Heaviside step function)
\begin{eqnarray}\label{eq:Gret3} 
  G(\bm{k},t)&=&\mathrm{i}\Theta(t)\frac{c}{2k}\Big(
  \exp[-\,\mathrm{i}c(k-\mathrm{i}\epsilon)t]
  -\exp[\mathrm{i}c(k+\mathrm{i}\epsilon)t]\Big).
\end{eqnarray}
Substitution of Eqs.~(\ref{eq:Gret2},\ref{eq:Gret3}) into (\ref{eq:solvamp1}) yields
\begin{eqnarray}\nonumber
  \bm{\mathcal{A}}(\bm{x},t)&=&\mathrm{i}\frac{\mu_0 c}{\sqrt{2\pi}^3}
  \int\!\frac{d^3\bm{k}}{2k}\int_{-\infty}^t\!dt'\,\Big(
  \exp[\mathrm{i}(\bm{k}\cdot\bm{x}-ckt)]
  \\ \nonumber
   &&\times \exp[\mathrm{i}c(k-\mathrm{i}\epsilon)t']
    \bm{j}_\perp(\bm{k},t')
    \\ \label{eq:solvamp2}
    &&-\exp[-\,\mathrm{i}(\bm{k}\cdot\bm{x}-ckt)]
    \exp[-\,\mathrm{i}c(k+\mathrm{i}\epsilon)t']
    \bm{j}^+_\perp(\bm{k},t')
  \Big).
\end{eqnarray}
Here we used $\bm{j}_\perp(-\,\bm{k},t)=\bm{j}^+_\perp(\bm{k},t)$.

We can therefore produce (\ref{eq:solvamp2}) as an expectation value (\ref{eq:solvamp1})
of the interaction picture operator $\bm{A}(\bm{x},t)$
if we use
\begin{eqnarray} \label{eq:xij1} 
  \bm{\xi}(\bm{k},t)&=&\mathrm{i}\sqrt{\frac{\mu_0 c}{2\hbar k}}
  \int_{-\infty}^t\!dt'\,\exp[\mathrm{i}c(k-\mathrm{i}\epsilon)t']
  \bm{j}_\perp(\bm{k},t')
\end{eqnarray}
in Eq.~(\ref{eq:xistate1}).

The liberty to add a constant amplitude $\bm{\zeta}(\bm{k})$
to $\bm{\xi}(\bm{k},t)$ corresponds to the possibility to add
an arbitrary solution of the homogenous wave equation to the
solution (\ref{eq:solvamp2}) of Amp\`{e}re's law (\ref{eq:ampere}).
The virtue of Eq.~(\ref{eq:xij1}) is to relate the photon amplitudes $\bm{\xi}(\bm{k},t)$
to the generating current density. In the representation (\ref{eq:fixxi1}), the photon
modes $\exp[\mathrm{i}(\bm{k}\cdot\bm{x}-ckt)]$ still oscillate with frequency $ck$
and move with velocity $c\hat{\bm{k}}$, while the generating current adjusts 
the amplitudes $\bm{\xi}(\bm{k},t)$ of the photons such that the observed
electromagnetic fields are created.

To ensure that the coherent
state also satisfies the interaction picture evolution equation
with the interaction picture Hamiltonian
\begin{eqnarray}\nonumber
  V_D(t)&=&-\int\!d^3\bm{x}\,\bm{j}_\perp(\bm{x},t)\cdot\bm{A}(\bm{x},t)
  \\ \nonumber
   &=&-\int\!d^3\bm{k}\,\sqrt{\frac{\hbar\mu_0c}{2k}}\Big[\exp(-\,\mathrm{i}ckt)
     \bm{a}(\bm{k})\cdot\bm{j}_\perp(-\,\bm{k},t)
     \\ \label{eq:HxiDt}
     &&+\,\exp(\mathrm{i}ckt)
     \bm{a}^+(\bm{k})\cdot\bm{j}_\perp(\bm{k},t)\Big],
\end{eqnarray}
i.e.
 \begin{eqnarray}\nonumber
   \mathrm{i}\hbar\frac{d}{dt}\bm{|}\bm{\xi}_D(t)\bm{\rangle}&=&
   V_D(t)\bm{|}\bm{\xi}_D(t)\bm{\rangle},
 \end{eqnarray}
 we can add a phase factor 
 \begin{equation}\label{eq:xiDt}
\bm{|}\bm{\xi}_D(t)\bm{\rangle}=\bm{|}\bm{\xi}(t)\bm{\rangle}\exp[-\,\mathrm{i}\Phi(t)/\hbar],
 \end{equation}  
 where
 \begin{eqnarray}\nonumber
   \frac{d}{dt}\Phi(t)&=&\mathrm{i}\int_{-\infty}^t\!dt'\int\!d^3\bm{k}\,\frac{\mu_0c}{4k}
   \Big(\exp[-\,\mathrm{i}c(k+\mathrm{i}\epsilon)t']\bm{j}_\perp(-\,\bm{k},t')
   \cdot\bm{j}_\perp(\bm{k},t)
   \\ \nonumber
   &&\times\exp[\mathrm{i}c(k-\mathrm{i}\epsilon)t]
     \\ \label{eq:PhasePhit}
     &&-\,\exp[-\,\mathrm{i}c(k+\mathrm{i}\epsilon)t]
     \bm{j}_\perp(-\,\bm{k},t)
     \cdot\bm{j}_\perp(\bm{k},t')\exp[\mathrm{i}c(k-\mathrm{i}\epsilon)t']\Big).
 \end{eqnarray}
 Of course, this does not change Eq.~(\ref{eq:fixxi1}),
\begin{eqnarray}  \nonumber
  \bm{\mathcal{A}}(\bm{x},t)&=&\bm{\langle}\bm{\xi}_D(t)\bm{|}\bm{A}(\bm{x},t)
  \bm{|}\bm{\xi}_D(t)\bm{\rangle}
  \\ \nonumber
  &=&\sqrt{\frac{\hbar\mu_0 c}{(2\pi)^3}}
\int\!\frac{d^3\bm{k}}{\sqrt{2k}}\,
\Big(\bm{\xi}(\bm{k},t)
\exp[\mathrm{i}(\bm{k}\cdot\bm{x}-ckt)]
\\ \label{eq:fixxi1D}
&&+\,\bm{\xi}^+(\bm{k},t)
\exp[-\,\mathrm{i}(\bm{k}\cdot\bm{x}-ckt)]\Big),
\end{eqnarray}

Eqs.~(\ref{eq:xistate1},\ref{eq:xij1},\ref{eq:xiDt},\ref{eq:PhasePhit}) and (\ref{eq:fixxi1D})
yield the
photon decomposition of electromagnetic fields
generated by the current $\bm{j}_\perp(\bm{x},t)$ in the form
\begin{eqnarray}  \nonumber
  \bm{\mathcal{B}}(\bm{x},t)&=&\bm{\nabla}\times\bm{\mathcal{A}}(\bm{x},t)
  =\bm{\langle}\bm{\xi}_D(t)\bm{|}\bm{\nabla}\times\bm{A}(\bm{x},t)
  \bm{|}\bm{\xi}_D(t)\bm{\rangle}
  \\ \label{eq:Bxi1}
  &=&\bm{\langle}\bm{\xi}_D(t)\bm{|}\bm{B}(\bm{x},t)
  \bm{|}\bm{\xi}_D(t)\bm{\rangle}
\end{eqnarray}
and
 \begin{eqnarray}  \nonumber
   \bm{\mathcal{E}}_\perp(\bm{x},t)&=&-\,\frac{\partial\bm{\mathcal{A}}(\bm{x},t)}{\partial t}
  =-\,\bm{\langle}\bm{\xi}_D(t)\bm{|}\frac{\partial\bm{A}(\bm{x},t)}{\partial t}
  \bm{|}\bm{\xi}_D(t)\bm{\rangle}
  \\ \label{eq:Exi1}
  &=&\bm{\langle}\bm{\xi}_D(t)\bm{|}\bm{E}_\perp(\bm{x},t)
  \bm{|}\bm{\xi}_D(t)\bm{\rangle}.
 \end{eqnarray}
 
 In Eq.~(\ref{eq:Exi1}), we used that the interaction picture Hamiltonian
 (\ref{eq:HxiDt}) satisfies\footnote{Eq.~(\ref{eq:commHDA}) holds for any photon-matter coupling
   Hamiltonian that, in the Schr\"odinger picture and in Coulomb gauge, only
   contains $\bm{A}(\bm{x})$
   as a photon operator. As a consequence, Eq.~(\ref{eq:Exi1}) will also hold in these models.}
 \begin{eqnarray} \label{eq:commHDA}
   [V_D(t),\bm{A}(\bm{x},t)]&=&0.
 \end{eqnarray}

\section{Photonic resolution of static magnetic fields}\label{sec:magnet}

Eq.~(\ref{eq:xij1}) yields for stationary current the photon amplitude
\begin{eqnarray} \label{eq:xicirc1}
  \bm{\xi}(\bm{k},t)&=&\sqrt{\frac{\mu_0}{2\hbar ck}}\frac{\exp(\mathrm{i}ckt)}{k-\mathrm{i}\epsilon}
  \bm{j}_\perp(\bm{k})
  =\bm{\xi}(\bm{k})\exp(\mathrm{i}ckt),
\end{eqnarray}
while the phase from (\ref{eq:PhasePhit}) becomes
\begin{equation}\label{eq:PhiisEt}
\Phi(t)=-\int\!d^3\bm{k}\,\frac{\mu_0}{2k^2}|\bm{j}(\bm{k})|^2t,
\end{equation}
which we can understand to be
\begin{equation}\label{eq:PhiisEt2}
\Phi(t)=Et.
\end{equation}
The time-dependence in Eq.~(\ref{eq:xicirc1}) and the results
(\ref{eq:PhiisEt},\ref{eq:PhiisEt2}) make sense, because
\begin{equation}
  \bm{|}\bm{\xi}_D(t)\bm{\rangle}=\bm{|}\bm{\xi}(t)\bm{\rangle}\exp(-\,\mathrm{i}Et/\hbar),
\end{equation}
with $\bm{\xi}(\bm{k},t)$ given by (\ref{eq:xicirc1}), 
is exactly the Dirac-picture representation of a stationary
state $\bm{|}\bm{\xi}_S(t)\bm{\rangle}$ in the Schr\"odinger picture\footnote{The transformation
  (\ref{eq:xiS1}) into the Dirac picture uses
  $\exp(\mathrm{i}H_\gamma t/\hbar)a^+(\bm{k})\exp(-\,\mathrm{i}H_\gamma t/\hbar)
  =a^+(\bm{k})\exp(\mathrm{i}ckt)$.},
\begin{equation}
\bm{|}\bm{\xi}_S(t)\bm{\rangle}=\bm{|}\bm{\xi}\bm{\rangle}
  \exp(-\,\mathrm{i}Et/\hbar),
  \end{equation}
\begin{equation}\label{eq:xiS1}
  \bm{|}\bm{\xi}_D(t)\bm{\rangle}=\exp(\mathrm{i}H_\gamma t/\hbar)\bm{|}\bm{\xi}\bm{\rangle}
  \exp(-\,\mathrm{i}Et/\hbar).
\end{equation}
We expect a constant electromagnetic field
to correspond to a stationary state in the Schr\"odinger picture.
We can confirm Eqs.~(\ref{eq:PhiisEt},\ref{eq:PhiisEt2}) with the Hamiltonian
\begin{equation}
  H=H_\gamma+V.
\end{equation}
Here, $H_\gamma$ is given in Eq.~(\ref{eq:Hgamma}) and the Schr\"odinger picture
coupling term is
\begin{eqnarray}\nonumber
  V&=&-\int\!d^3\bm{x}\,\bm{A}(\bm{x})\cdot\bm{j}(\bm{x})
  \\ \nonumber
  &=&-\int\!d^3\bm{k}\,\sqrt{\frac{\hbar\mu_0 c}{2k}}\left[\bm{a}(\bm{k})\cdot\bm{j}^+(\bm{k})
    +\bm{a}^+(\bm{k})\cdot\bm{j}(\bm{k})\right]\!.
\end{eqnarray}
The state $\bm{|}\bm{\xi}\bm{\rangle}$, with $\bm{\xi}(\bm{k})$ given in Eq.~(\ref{eq:xicirc1}),
does indeed solve the time-independent Schr\"odinger equation
\begin{eqnarray}\nonumber
  H\bm{|}\bm{\xi}\bm{\rangle}&=&-\int\!d^3\bm{k}\,\frac{\mu_0}{2k^2}|\bm{j}(\bm{k})|^2
  \bm{|}\bm{\xi}\bm{\rangle}
  =-\int\!d^3\bm{k}\,\hbar ck|\bm{\xi}(\bm{k})|^2\bm{|}\bm{\xi}\bm{\rangle}
  \\
  &=&-\int\!d^3\bm{k}\,\hbar ck\,n(\bm{k})\bm{|}\bm{\xi}\bm{\rangle},
\end{eqnarray}
where $n(\bm{k})$ is the $\bm{k}$-space density of the photons.

The result for the energy eigenvalue,
\begin{equation}\label{eq:Evalue}
  E=\bm{\langle}\bm{\xi}\bm{|}H\bm{|}\bm{\xi}\bm{\rangle}
  =-\int\!d^3\bm{k}\,\hbar ck\,n(\bm{k})
  =-\,\bm{\langle}\bm{\xi}\bm{|}H_\gamma\bm{|}\bm{\xi}\bm{\rangle},
\end{equation}
might surprise at first. However, it agrees with a virial theorem for static
magnetic fields. If we define the electromagnetic virial
as $\int d^3\bm{x}\,\epsilon_0\bm{A}\cdot\dot{\bm{A}}/2$, we have
\begin{eqnarray} \nonumber
  &&\frac{\epsilon_0}{2}\frac{d}{dt}\int d^3\bm{x}\,\bm{A}(\bm{x},t)\cdot
  \dot{\bm{A}}(\bm{x},t)
  \\ \nonumber
  &&=\frac{\epsilon_0}{2}\int d^3\bm{x}\,\dot{\bm{A}}^2(\bm{x},t)
  +\frac{1}{2\mu_0}\int d^3\bm{x}\,\bm{A}(\bm{x},t)\cdot\left[
    \Delta\bm{A}(\bm{x},t)+\mu_0\bm{j}(\bm{x},t)\right]
  \\ 
  &&=\frac{\epsilon_0}{2}\int d^3\bm{x}\,\dot{\bm{A}}^2(\bm{x},t)
  +\frac{1}{2}\int d^3\bm{x}\left[
    \bm{A}(\bm{x},t)\cdot\bm{j}(\bm{x},t)-\frac{1}{\mu_0}\bm{B}^2(\bm{x},t)\right].
\end{eqnarray}
For static electromagnetic fields, this implies
\begin{equation}
  \langle V\rangle=-\,2\langle H_\gamma\rangle,
\end{equation}
in agreement with Eq.~(\ref{eq:Evalue}).

 For a simple application, we consider the magnetic field of a current $I$ that runs
in a thin circular wire of radius $R$ in the $(x,y)$-plane,
\begin{equation}\label{eq:jcirc1}
  \bm{j}(\bm{x})=I\delta(z)\delta\!\left(\sqrt{x^2+y^2}-R\right)\bm{e}_\varphi.
  \end{equation}
The vector
\begin{equation}
  \bm{e}_\varphi=\left(\begin{array}{c}
    -\,\sin\varphi\\ \,\,\,\,\,\,\cos\varphi\\ \,\,\,\,\,0\\
    \end{array}\right)
\end{equation}
is the normal vector in azimuthal direction in the $(x,y)$-plane.

The Fourier transformed current density is
\begin{equation}\label{eq:jcirck}
  \bm{j}(\bm{k})=-\,\mathrm{i}\frac{IR}{\sqrt{2\pi}}J_1(kR\sin\theta)\bm{e}_\phi,
\end{equation}
where $\theta$ and $\phi$ are polar and azimuthal angles in $\bm{k}$-space.

The current density (\ref{eq:jcirc1}) is already transversal, $\bm{\nabla}\cdot\bm{j}(\bm{x})=0$,
and Eq.~(\ref{eq:xicirc1}) yields the $\bm{k}$-space photon density for the magnetic field,
\begin{equation}\label{eq:ncirck}
  n(\bm{k})=\left|\bm{\xi}(\bm{k},t)\right|^2
  =\frac{\mu_0 I^2R^2}{4\pi\hbar ck^3}J_1^2(kR\sin\theta).
\end{equation}

Integration of $n(\bm{k})$ yields the total number of photons in the magnetic field,
\begin{equation}
N=\frac{\mu_0 I^2R^2}{2\hbar c}.
\end{equation}

\section{Conclusions}\label{sec:conc}

Eq.~(\ref{eq:xij1}) yields the photon amplitudes in coherent states
(\ref{eq:xistate1},\ref{eq:xiDt},\ref{eq:PhasePhit}) that describe
electromagnetic waves in terms of their source currents. The states
are Dirac-picture states, such that the corresponding electromagnetic fields are
given through the expectation values of the free photon quantum fields
(\ref{eq:Bxi1},\ref{eq:Exi1}). Furthermore, the states are not limited to the
far-field regime.

The motivation to derive these states was driven by a
conceptual desire to better understand the photon content of macroscopic
electromagnetic fields. However, 
Eqs.~(\ref{eq:xistate1},\ref{eq:xij1},\ref{eq:xiDt},\ref{eq:PhasePhit})
should also prove useful in mesoscopic
integrations of photonics in quantum technologies.

\subsection*{Acknowledgments}

We acknowledge support from the Natural Sciences and Engineering Research Council
of Canada.

\subsection*{Data Availability Statement}

No new data were created or analysed in this study.\\


\end{document}